\documentclass[iop]{emulateapj}
\usepackage{amsmath}
\usepackage{amssymb}
\usepackage{color}
\usepackage{resizegather}
\usepackage[normalem]{ulem}

\newcommand{\change}[1]{\textcolor{black}{#1}}
\newcommand{\secondchange}[1]{\textcolor{black}{#1}}

\newcommand{\fn}{\widehat{w}}

\shorttitle{Galaxy Redshifts from Discrete Optimization of Correlation Functions}
\shortauthors{Lee, Budav\'ari, Basu \& Rahman}

\begin{document}

\title{Galaxy Redshifts from Discrete Optimization of Correlation Functions}

\author{Benjamin C.G. Lee\altaffilmark{1}} 
\author{Tam\'as Budav\'ari\altaffilmark{2,3,4}}
\author{Amitabh Basu\altaffilmark{2}}
\author{Mubdi Rahman\altaffilmark{4}}

\altaffiltext{1}{Harvard College, Cambridge, MA 02138, USA}
\altaffiltext{2}{Dept.\ of Applied Mathematics \& Statistics, Johns Hopkins University, 3400 N. Charles St., Baltimore, MD 21218, USA}
\altaffiltext{3}{Dept.\ of Computer Science, Johns Hopkins University,  3400 N. Charles St., Baltimore, MD 21218, USA}
\altaffiltext{4}{Dept.\ of Physics \& Astronomy, Johns Hopkins University,  3400 N. Charles St., Baltimore, MD 21218, USA}

\email{Email: benjaminlee@college.harvard.edu}

\begin{abstract}

We propose a new method of \change{constraining} the redshifts of individual extragalactic sources based on \change{celestial coordinates and their ensemble statistics}. 
Techniques from integer linear programming are utilized to optimize simultaneously for the angular two-point cross- and autocorrelation functions.
Our novel formalism introduced here not only transforms the otherwise hopelessly expensive, brute-force combinatorial search into a linear system with integer constraints but also is readily implementable in off-the-shelf solvers. We adopt \change{Gurobi, a commercial optimization solver,} and use Python to build the cost function dynamically.
The preliminary results on simulated data show \change{potential} for future applications to sky surveys by complementing and enhancing photometric redshift estimators.
Our approach is the first \change{application of integer linear programming to astronomical analysis.} 
\end{abstract}


\section{Introduction}\label{sec:introduction}

Knowing the redshifts of extragalactic objects is vital for understanding their physical properties as well as for statistical cosmology measurements.
The primary method employed by researchers to measure galaxy redshifts is spectroscopy; however, spectroscopy is  available for only a small fraction of all imaged galaxies, on the order of 1\% \citep{rahmanb}. 
This low percentage has necessitated a search for alternative methods of redshift estimation.

Much progress has been made with photometric redshift estimation methods, which primarily rely on the colors of galaxies to estimate redshifts \citep{koo85, connolly95, koo99, benitez00, budavari00, budavari01, feldmann06, brammer08, budavari08}.  While such methods are used frequently, they face inherent limitations, resulting in redshift estimates with large associated error values.  Consequently, methods of redshift estimation using the angular clustering of galaxies have been proposed and explored \citep{benjamin10, menard13, newman08, rahmana, rahmanb, schmidt2013, schmidt2015}.  Because these methods require only the angular positions of galaxies, they are considered distinct from photometric redshift methods per the working definition argued by \citet{koo99}.

The ``clustering-based'' redshift estimation methods rely on the spatial correlations and directly use the angular two-point correlation function $w(\theta)$, which expresses the excess probability of finding a galaxy at an angular separation $\theta$ from another galaxy.  
While these methods are promising, they propose only redshift distributions and thus are agnostic to the redshifts of specific galaxies.  Therefore, a natural question arising from clustering-based redshift estimation is whether galaxy clustering can be used not only to produce redshift distributions of galaxy samples but also to determine the redshifts of individual galaxies.

In this paper, we propose a new method that uses combinatorial optimization techniques applied to the angular two-point correlation function to divide galaxy samples into separate redshift bins, thereby constraining the redshifts of individual galaxies. Creating thin slices not only is possible but potentially makes the optimization faster. Ultimately, the final method is expected to balance the statistical noise and the optimization cost.  %{\color{blue}{Indeed, using such a method first to constrain redshifts and then to determine cross-correlations with other datasets would require the target correlation function to be treated as a prior on the dataset.}}
\change{In practice the true correlation functions are not known, but accurate prior information is usually available. In this paper, we first study the ideal case and then move on to testing the limits of the approximation.}

Here we explore the simplest case. We aim to partition a set of (simulated) galaxies into two subsamples such that their correlation functions match the observations.  This case would be particularly useful in creating a hard boundary between overlapping photo-z redshift bins.  If it works, the optimization can be repeated iteratively in order to separate a galaxy sample into narrow redshift slices.  
Of course, even this simplest case presents computational challenges.
\change{Naively, for a sample of size $n$ and fixed subsample sizes of $k$ and \mbox{$n\!-\!k$}, $n$ choose $k$ possible partitions would have to be considered in order to find the optimal partition, which is prohibitively expensive.}
\iffalse
\sout{Naively, for a sample of size $n$, $2^n$ possible partitions would have to be considered in order to find the optimal partition, which is prohibitively expensive even for dozens of objects.}
\fi
We present a formalism using integer linear programming that, when implemented, makes this optimization tractable for statistically relevant sample sizes, \change{given the  appropriate conditions}.
\change{We then test this method using mock catalogs and Gurobi, an optimization solver, as a proof of concept.  In order to restrict our analysis to the  mathematical effectiveness of the optimization itself, we use the ground-truth correlation functions of the samples as the target values of the optimization, thereby eliminating physical considerations from the method.}

In Section \ref{sec:formalism}, we provide a detailed description of our method, including a derivation of the relevant formalism.  In Section \ref{sec:results}, we present results from \change{our implementation of this method.} In Section \ref{sec:applicability}, we discuss the applicability of this method.

\section{Our Approach}\label{sec:formalism}

Before introducing our formalism, let us introduce the notation that will be used throughout the rest of this paper.  Let $D$ and $D'$ be two datasets with the same sky coverage, and let $R$ be a random dataset with the same sky coverage as $D$ and $D'$.    In accordance with \citet{landy} we define $DD$ as the set of all unordered pairs of galaxies in $D$, and we define $DD'$ as the set of all unordered pairs of galaxies such that one galaxy is from $D$ and the other is from $D'$.  We also define $D'D', DR, D'R,$ and $RR$ analogously. Because the correlation function is estimated over a set of $n$ bins (i.e., intervals) of angular separation, we must introduce notation related to pair counts within these bins.  We define $DD_i$ as the set of unordered data-data pairs such that the angular separation between the members of the pair is in bin $i$, $i = 1, \ldots, n$.  The corresponding terms for $DR$, $RR$, etc., are defined analogously.  Lastly, we define $|D|$ as the size of $D$, $|DD_i|$ as the size of $DD_i$, etc.

In this paper, we develop our formalism using two estimators of correlation functions given two samples with the same sky coverage: the natural estimator and the \citet{landy} estimator.  The natural estimator is the least expensive of the  correlation function estimators to compute and hence expected to yield the simplest optimization formulas. On the other hand, the Landy-Szalay estimator has been shown to be the most accurate estimator by \citet{kerscher}. 
Using the natural estimator, the cross-correlation and autocorrelation function estimates of the two samples $D$ and $D'$ in bin $i$ are given by:
\begin{equation}\label{eq:daffy-duck}
\fn_{X, i} = \frac{|R|(|R| - 1)}{2 \cdot |RR_i|} \cdot \frac{|DD'_i|}{|D||D'|} - 1
\end{equation}
\begin{equation}
\fn_{D, i} = \frac{|R|(|R| - 1)}{|RR_i|} \cdot \frac{|DD_i|}{|D|(|D| - 1)} - 1
\end{equation}
\begin{equation}
\fn_{D', i} = \frac{|R|(|R| - 1)}{|RR_i|} \cdot \frac{|D'D'_i|}{|D'|(|D'| - 1)} - 1
\end{equation}
In the main body of the paper, we focus on these equations but provide the alternative method for the Landy-Szalay estimator in Appendix~\ref{sec:LS}.

\subsection{Integer Linear Programming (ILP)}\label{sec:ILP} 

One of the most powerful tools in discrete optimization is {\em integer linear programming (ILP)}. {\em Linear Programming} is an optimization model in which one seeks to optimize a linear function of finitely many variables, subject to linear inequality constraints on the variables. Integer Linear Programming generalizes this by only admitting solutions where some (or all) of the variables are constrained to take only integer values. Such models are an indispensable tool for optimizing over discrete solution spaces, and state-of-the-art software codes have been developed to handle numerous large scale problems arising in technological, scientific, and economic applications.

In order to build an integer linear programming model, we must build a model consisting of a cost function and constraints, where our cost function and constraints are constructed such that the minimization of our cost function yields the optimal solution to our problem.  We define our optimal solution as a partition of our sample $V$ into two subsamples $S$ and its complement $\overline{S}$ such that:
\begin{equation}\label{eq:tot}
\sum_{i = 1}^{n} \bigg[ \big|\fn_{S, i} - \alpha_i \big| + \big|\fn_{\overline{S}, i} - \beta_i \big| + \big|\fn_{X, i} - \gamma_i \big|    \bigg]
\end{equation}
is minimized, where $\fn_{S, i}$, $\fn_{\overline{S}, i}$, and $\fn_{X, i}$ are our correlation function estimates in bin $i$, as calculated using our choice of estimator, and $\alpha_i, \beta_i,$ and $\gamma_i$ are target values for $\fn_{S, i}$, $\fn_{\overline{S}, i}$, and $\fn_{X, i}$, respectively.    Informally, this function is minimized when $\fn_{S}$, $\fn_{\overline{S}}$, and $\fn_{X}$ are pulled as close to our target values $\alpha_i, \beta_i,$ and $\gamma_i$ as possible, across all bins $i$.  

The purpose of the following formalism is to translate equation \ref{eq:tot} into an integer linear programming model.  The construction of such a model requires translating all unknowns into variables, creating a cost function using a linear combination of these variables, and adding constraints to the model to enforce certain relationships between the variables. We must also specify which variables are integer or continuous. In current software, one can even specify if an integer variable is binary, i.e., takes only values $0$ or $1$. Our integer variables will, in fact, be binary variables to accommodate the problem of classifying objects into two redshift slices.

To simplify the optimization further, we fix the size of $S$ and its complement, $|S|$ and $|\overline{S}|$, to prevent runaway solutions in which one subsample contains a large majority of the galaxies in $V$. Moreover, this enables us to keep the cost function and constraints linear.  Consequently, we treat $|S|$ and $|\overline{S}|$ as constants throughout. This, however, poses no real threat to generality because often good initial estimates are available, and further optimization can be performed along the size dimension if needed.

We begin by defining variables for each of our galaxies \mbox{$u\!\in\!V$}.  We introduce the binary variable $x_u$ that encodes whether a galaxy $u$ is a member of $S$ or $\overline{S}$:
\begin{equation}\label{eq:x-definition}
   x_u = \left\{
     \begin{array}{lr}
       1 & : u \in S\\
       0 & : u \in \overline{S}
     \end{array}
   \right.
\end{equation} 
These variables serve as the bridge between the cost function and the partitioning of $V$: we will construct the cost function in such a way that minimizing it sets each $x_u$ to either 0 or 1 and thus assigns each galaxy to $S$ or $\overline{S}$ according to the optimal partition.

We now add a constraint to our model in order to enforce that $|S|$ must be fixed to a pre-determined positive integer by using the fact that $|S|$ is precisely the sum of $x_u's$ that evaluate to $1$:
\begin{equation}\label{eq:ssize}
\sum_{u \in V} x_u = |S|
\end{equation}
Thus, we have fixed $|S|$, and because \mbox{$V\!=\!S\!\sqcup\!\overline{S}$}, where $|V|$ is fixed, we have also fixed $|\overline{S}|$.

Next, we introduce variables for unordered pairs of galaxies in $V$ that encode whether the galaxies in each pair are from the different subsamples.  For each unordered pair of galaxies \mbox{$(u, v)\!\in\!VV$}, we define the binary variable $y_{uv}$ as follows:
\begin{equation}\label{eq:y-definition}
   y_{uv} = \left\{
     \begin{array}{lr}
       1 & : u, v \text{ are in different subsamples}\\
       0 & : u, v \text{ are in the same subsample}
     \end{array}
   \right.
\end{equation} 
where $y_{uv}$ is symmetric in $u$ and $v$.  Significantly, $y_{uv}$ can be expressed in terms of the Boolean \mbox{\textit{``exclusive or''}} (XOR hereafter) of the $x_u$ and $x_v$ variables, 
\begin{equation}
y_{uv} = x_u \oplus x_v
\end{equation}
where each XOR is encoded through four linear constraints of $x_u, x_v,$ and $y_{uv}$ that we add to our model as
\begin{equation}
  \begin{split}
  y_{uv} \ge x_u - x_v
  \qquad
&y_{uv} \ge x_v - x_u \\
y_{uv} \le x_u + x_v
\qquad
&y_{uv} \le 2 - x_u - x_v
  \end{split}
\end{equation}
We reiterate that these constraints establish the relationship between $x$'s and $y$'s, as given in their definitions in \eqref{eq:x-definition} and \eqref{eq:y-definition}.

Before proceeding, we introduce more notation-related summations over pair counts.  We use the summation notation $\sum\limits_{(u, v) \in VV_i}$ to denote summing over all unordered pairs of galaxies in $V$ such that the angular separation between $u$ and $v$ falls into bin $i$.

We begin by translating the natural estimator $\fn_N$ to its linear programming equivalent in this section and generalize to the Landy-Szalay estimator $\fn_{LS}$ in the Appendix.

\subsection{Cross-correlation Function}\label{sec:cross}
Using the natural estimator, we first translate $\fn_{X}$ into its cost function equivalent, $f_{X}(S)$.  For cross-correlation, we seek to minimize:
\begin{equation}
\sum_{i = 1}^{n} \bigg|\fn_{X, i} - \gamma_i \bigg|
\end{equation}
where $\fn_{X, i}$ is an estimate of cross-correlation in bin $i$.
We can express $\fn_{X, i}$ in terms of previously-defined quantities:
\begin{equation}
\fn_{X, i} = \frac{|R|(|R| - 1)}{2 \cdot |RR_i|} \cdot \frac{|S\overline{S}_i|}{|S||\overline{S}|} - 1
\end{equation}
Because $R$ is fixed, $|R|$ and $|RR_i|$ are constants.  Furthermore, $|S|$ and $|\overline{S}|$ are fixed.  Thus, we can combine these constants into a single weight for each bin:
\begin{equation}\label{eq:a}
a_i =  \frac{|R|(|R| - 1)}{2 \cdot |RR_i|} \cdot \frac{1}{|S||\overline{S}|}
\end{equation}

We therefore seek to minimize
\begin{equation}
\sum_{i = 1}^{n} \bigg|a_i |S \overline{S}_i| - (1 + \gamma_i) \bigg|
\end{equation}
where $|S \overline{S}_i|$ is the only non-constant term within the minimization for each bin $i$.  We can now reformulate this expression using our previously-defined binary variables $y_{uv}$. $|S\overline{S}_i|$ is precisely equal to the number of unordered pairs $u, v$ of galaxies in $V$ such that $u$ and $v$ are in different subsamples and the angular separation between $u$ and $v$ falls into bin $i$; thus,
\begin{equation}
 |S\overline{S}_i| = \sum_{(u, v) \in VV_i} y_{u v} 
\end{equation}

Because we are summing over all unordered pairs in $VV_i$, as opposed to just unordered pairs in $S \overline{S}_i$, we have eliminated any dependence on the partitioning of $V$ except in the variables themselves.  
The expression that we seek to minimize for cross-correlation optimization now simplifies to:
\begin{equation}\label{eq:cross-corr}
\sum_{i = 1}^n \bigg|\bigg( a_i \sum_{(u, v) \in VV_i} y_{u v} \bigg) - (1 + \gamma_i) \bigg|
\end{equation}
Because each $y_{uv}$ is the XOR of $x_u$ and $x_v$, we have expressed cross-correlation optimization entirely in terms of binary variables for each galaxy, and minimization of the above expression will assign each galaxy to either $S$ or $\overline{S}$ according to the optimal partition of $V$.   

The expression in \eqref{eq:cross-corr} is not a linear function of the $y_{uv}$ variables due to the absolute value function. However, this expression can be modeled by a linear function by introducing auxiliary continuous variables $\varphi_i$ for each bin $i$ and relating them to the $y_{uv}$ variables as follows. 
\begin{equation}\label{eq:aux-var}
f_{X}(S) = \sum_{i = 1}^n \varphi_i
\end{equation}
where for each $\varphi_i$, we add the following two constraints:
\begin{equation}\label{eq:abs-val}
\begin{split}
\bigg( a_i \sum_{(u, v) \in VV_i} y_{u v} \bigg) - (1 + \gamma_i) &\le \varphi_i\\
\bigg( a_i \sum_{(u, v) \in VV_i} y_{u v} \bigg) - (1 + \gamma_i) &\le - \varphi_i
  \end{split}
\end{equation}
  We have now incorporated cross-correlation optimization into our linear programming model. The key insight here is that any solution that minimizes the expression in~\eqref{eq:aux-var} must satisfy one of the inequalities in~\eqref{eq:abs-val} at equality depending on which left hand side is larger, and thus 
  \begin{equation}
  \varphi_i = \bigg|\bigg( a_i \sum_{(u, v) \in VV_i} y_{u v} \bigg) - (1 + \gamma_i)\bigg|
  \end{equation}
  in any optimum solution minimizing~\eqref{eq:aux-var}.  We mention here the conscious choice of using the $L^1$ norm in~\eqref{eq:tot}, as opposed to the $L^2$ norm:  the $L^1$ norm leads to a formulation with a linear objective like~\eqref{eq:aux-var} and linear constraints like \eqref{eq:abs-val}, as opposed to the $L^2$ norm which would give a convex quadratic objective. As per folk wisdom in integer programming, we prefer the linear formulation, and hence we go with the $L^1$ norm.

Next, we formalize autocorrelation optimization using the natural estimator.  In this paper, we formalize two different approaches: combining the autocorrelations of $S$ and $\overline{S}$ in such a way that the autocorrelation target values $\alpha_i$ and $\beta_i$ of $S$ and $\overline{S}$, respectively, are set to be equal, and implementing separate target values for the autocorrelation of $S$ and $\overline{S}$, thereby allowing $\alpha_i$ and $\beta_i$ to be potentially distinct for any bin $i$. Although the latter method is advantageous because it allows for independent target values, the downside is that one needs more variables in the integer linear program to model this, compared to the former method in which no new variables need to be introduced into our model.  We introduce the former method next and introduce the latter method in the Appendix in \ref{sec:independent}.

\subsection{Autocorrelation Functions}\label{sec:combined}%, Combined Autocorrelation Method}
To derive a parameterized autocorrelation for the two samples, let us consider the variable $z_{uv}$ defined as 
\begin{equation}
%\begin{displaymath}
%  1 - y_{uv} = 
  z_{uv} = \left\{
     \begin{array}{lr}
       1 & : u,v \text{ are in the same subsample}\\
       0 & : u,v \text{ are in separate subsamples}
     \end{array} 
   \right.
%\end{displaymath}
\end{equation}
From the definition we can see that \mbox{$z_{uv}=1\!-\!y_{uv}$}.
Although $z_{uv}$ is agnostic as to which sample $u$ and $v$ belong, it does encode whether the pair contributes to an autocorrelation calculation.  We can naturally extend the notion of autocorrelation for $S$ and $\overline{S}$ into a combined autocorrelation given by:
\begin{equation}
\fn_{A, i} = \frac{|R|(|R| - 1)}{|RR_i|} \cdot \frac{|S S_i| + |\overline{S} \overline{S}_i|}{|S|(|S| - 1) + |\overline{S}|(|\overline{S}| - 1)}  - 1
\end{equation}
where $\fn_{A, i}$ is the weighted average of $\fn_{S, i}$ and $\fn_{\overline{S}, i}$:
\begin{equation}
\fn_{A, i} = \lambda \fn_{S, i} + \big(1 - \lambda \big) \fn_{\overline{S}, i}
\end{equation}
with
\begin{equation}
\lambda = \frac{|S|(|S| - 1)}{|S|(|S| - 1) + |\overline{S}|(|\overline{S}| - 1)}
\end{equation}
In this combined autocorrelation model, we seek to minimize
\begin{equation}
\sum_{i = 1}^{n} \bigg|\fn_{A, i} - \alpha_i \bigg|
\end{equation}
where $\alpha_i$ is the target value for the combined autocorrelation in bin $i$.
We introduce the weight $b_i$ to replace constants in $\fn_{A,i}$:
\begin{equation}
b_i= \frac{|R|(|R| - 1)}{|RR_i|} \cdot \frac{1}{|S|(|S| - 1) + |\overline{S}|(|\overline{S}| - 1)}
\end{equation}

Furthermore, we can express \mbox{$|S S_i|\!+\!|\overline{S} \overline{S}_i|$} entirely in terms of variables that have already been introduced because this sum is precisely equal to the sum of unordered pairs $(u, v)$ of galaxies in $V$ such that $u$ and $v$ are in the same subsample and the angular separation between $u$ and $v$ falls into bin $i$:
\begin{equation}
 |SS_i| + |\overline{S} \overline{S}_i| = \sum_{(u, v) \in VV_i} z_{u v}  = \sum_{(u, v) \in VV_i} (1 - y_{uv})
\end{equation}
Our expression takes the form:
\begin{equation}
\sum_{i = 1}^n  \bigg| \bigg[ b_i \sum_{(u, v) \in VV_i} (1 - y_{uv}) \bigg] - (1 + \alpha_i) \bigg|
\end{equation}

As with cross-correlation optimization, the final step is to eliminate absolute values from the cost function.  For each bin $i$, we add a continuous variable $\psi_i$.  The portion of the cost function $f_{A}(S)$ corresponding to combined autocorrelation optimization takes its final form:
\begin{equation}
f_{A}(S) = \sum_{i = 1}^n \psi_i
\end{equation}
where for each $\psi_i$, we add the following two constraints:
\begin{equation}
\begin{split}
\bigg[ b_i \sum_{(u, v) \in VV_i} (1 - y_{uv}) \bigg] - (1 + \alpha_i) &\le \psi_i\\
 \bigg[ b_i \sum_{(u, v) \in VV_i} (1 - y_{uv}) \bigg] - (1 + \alpha_i) &\le - \psi_i
  \end{split}
\end{equation}
We can now express our entire model using combined autocorrelation and the natural estimator.  The model consists of the cost function:
\begin{equation}
f(S) = f_{X}(S) + f_{A}(S)
\end{equation}
and all associated constraints.

This model can be generalized to the Landy-Szalay estimator by modifying the cost function slightly; we provide the details in the Appendix in \ref{sec:LS}.  Furthermore, in all of the above formalism, we have used a uniform weighting across all bins for cross-correlation and autocorrelation in the cost function for notational simplicity.  We note that one can easily weight each bin individually.

\begin{figure*}
\epsscale{1.17}
\plottwo{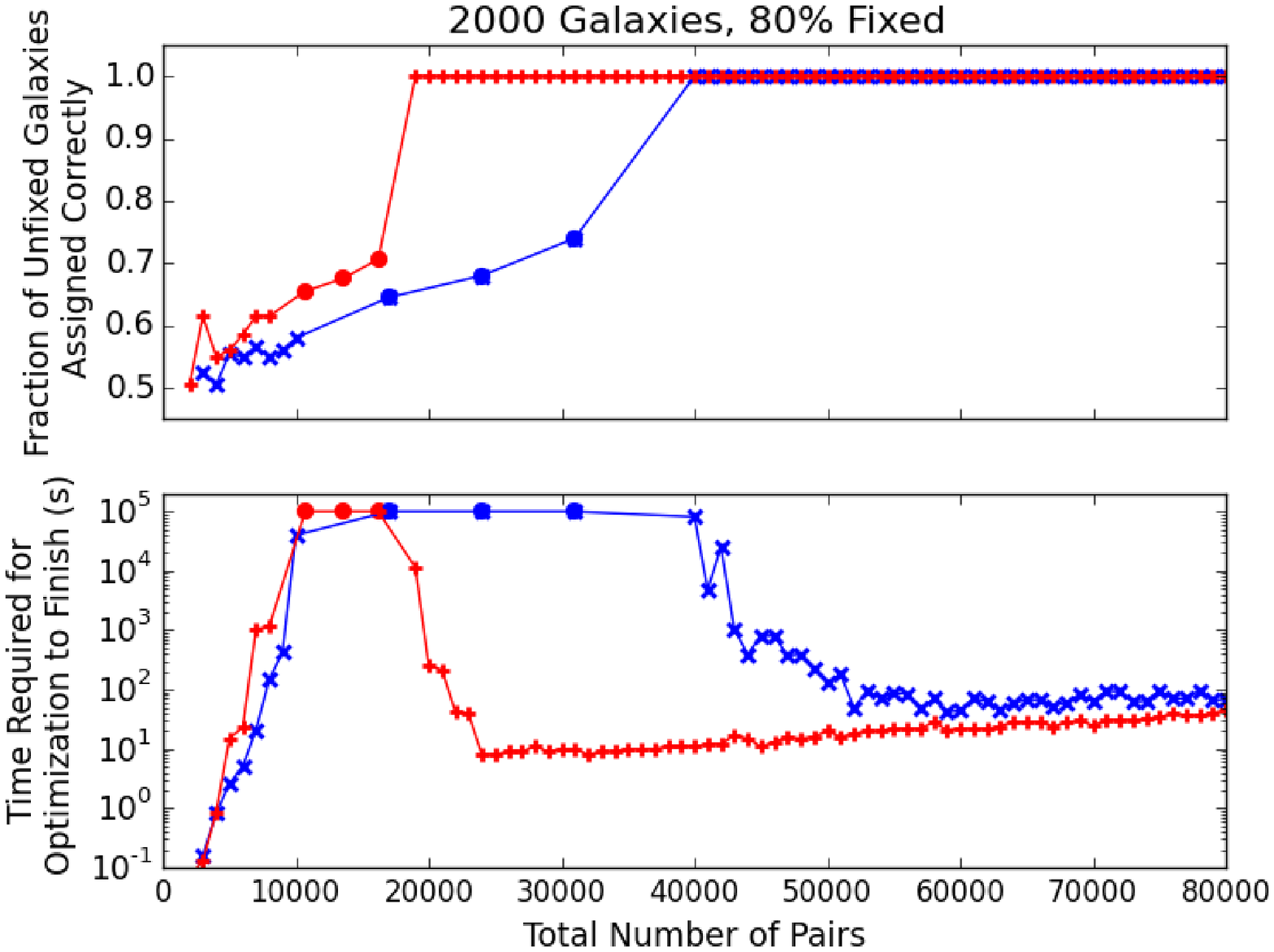}{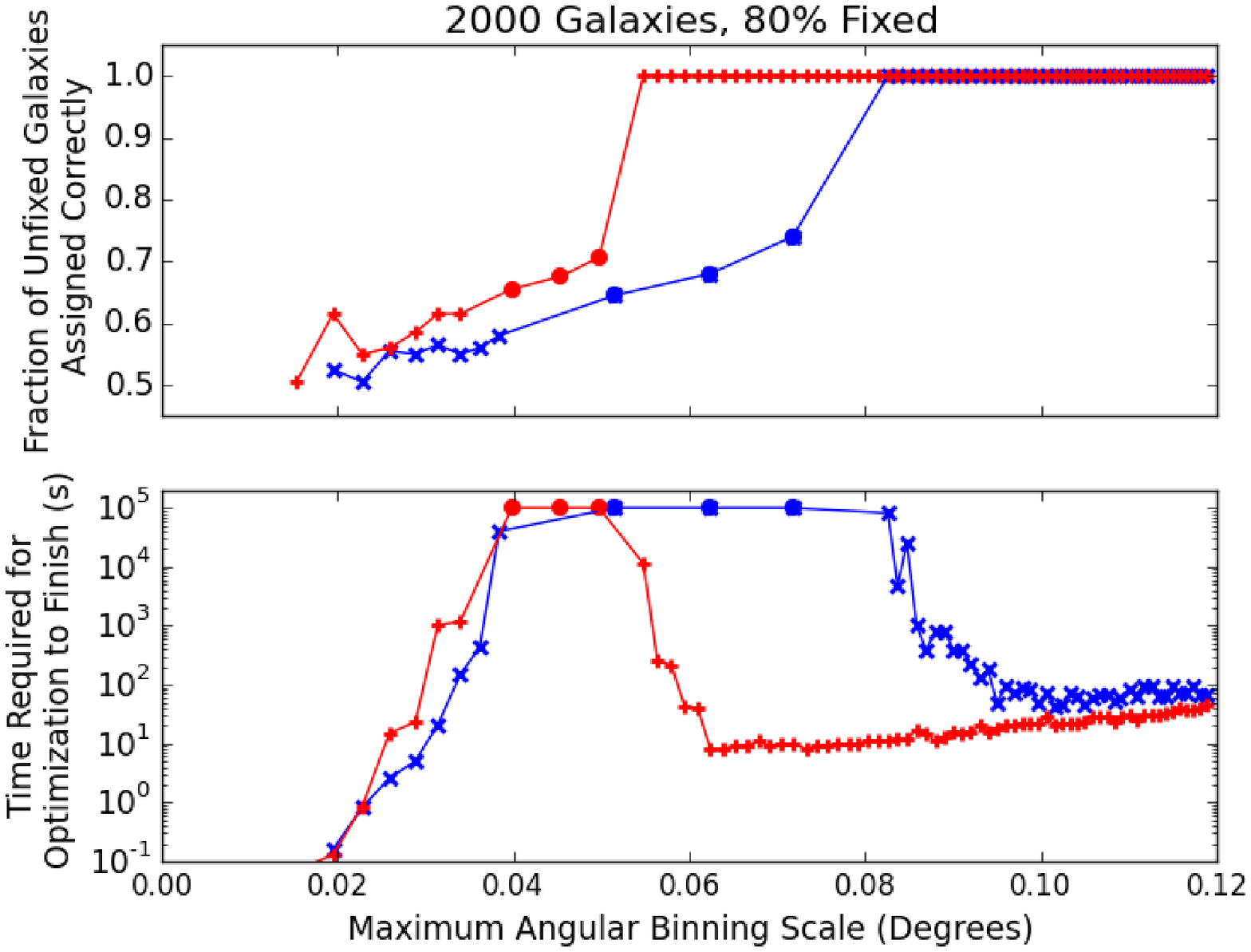}

    \caption{Plots of the fraction of unfixed galaxies assigned correctly and the time required for the optimization to finish as a function of number of bins  for a series of optimization runs with 2000 galaxies in $V$, 80\% of which were fixed.  \change{In the above plots, the red `+' markers correspond to runs using a binning scheme of 100 pairs per bin, and the blue `x' markers correspond to runs using a binning scheme of 200 pairs per bin.  The circles represent optimization runs that had not completed after $10^5$ seconds and were terminated after this amount of time; the reported fraction of unfixed galaxies assigned correctly for each of these runs was calculated using the optimal solution found before termination.} }
    
    \label{fig:100leftright}
\end{figure*}

\iffalse
\begin{figure*}
\epsscale{1.17}
\plottwo{f1a.eps}{f1b.eps}

    \caption{Plots of the fraction of unfixed galaxies assigned correctly and the time required for the optimization to finish as a function of number of bins  for a series of optimization runs with 2000 galaxies in $V$, 80\% of which were fixed, and a binning scheme of 100 pairs per bin.  {\color{blue}{In the above plots, the  circles}} correspond to optimization runs that completed in under $10^5$ seconds. The {\color{blue}{x's}} correspond to optimization runs that had not completed after $10^5$ seconds and were terminated after this amount of time; the reported fraction of unfixed galaxies assigned correctly was calculated using the optimal solution found before termination.  }
    
    \label{fig:100leftright}
\end{figure*}
\fi

\section{Implementation and Results}\label{sec:results}

To test the effectiveness of the integer linear programming method described in Section \ref{sec:formalism}, we used the Python interface for Gurobi,
 a solver available for both academic and commercial use.  We ran all tests on a Dell PowerEdge R815 machine with 512 GB of RAM and 4 AMD Opteron 6272 processors.  \change{We ran each core at a clock speed of $1200$ MHz, and the maximum memory bandwidth of the machine was 42.7 GB/s.}  The machine ran Scientific Linux 7, Python 2.7.5, and Gurobi 6.0.4.   It is worth noting that none of the following tests required Gurobi to use more than 1\% of RAM at any given time; therefore, our machine's specifications far exceed the specifications necessary to reproduce the following results.  \change{We allowed Gurobi to use all available cores, the default setting.  However, Gurobi uses a branch and bound algorithm, which requires that only a single core be used in the search of the root node, the first step in the algorithm; a more substantive discussion of this can be found on Gurobi's tutorial on mixed integer programming basics.\footnote{\change{http://www.gurobi.com/resources/getting-started/mip-basics}} Because the ILP problem in question requires a significant, but varying, amount of time to be devoted to the root node search, it is difficult to deduce a general, quantitative claim about Gurobi's runtime as a function of number of available cores for this application.\footnote{\change{More generally, research on parallel algorithms for integer linear programming is currently in its infancy. In the most powerful algorithms for integer linear programming, certain subroutines can be easily parallelized, while other components are inherently sequential in nature. Consequently, current commercial solvers like Gurobi implement ad-hoc techniques for exploiting multiple cores, which give highly varying results in practice. Designing parallel algorithms to solve integer linear programming is a topic of active research for the community, and breakthroughs are expected in the next 5-10 years.}} }

For the two datasets $S$ and $\overline{S}$, we used two mock catalogs, each 1 square degree in size and each consisting of 10,000 galaxies, generated 
\iffalse
\sout{by S\'ebastien Heinis}
\fi 
using a Cox point process, as described in Heinis et al. (2009).
The samples were produced using different random seeds, and consequently, they behaved as uncorrelated samples, meaning that the theoretical cross-correlation was 0.  In both samples, we selected the thin redshift cut of \mbox{$0.34\!\le\!z\!\le\!0.36$}, the densest redshift slice in both samples, in order to select samples with strong angular autocorrelation signals.  These cuts left slightly over 1000 galaxies per catalog.  We then randomly selected 2000 galaxies in total from these two samples, yielding $S$ and $\overline{S}$ such that \mbox{$|V|\!=\!2000$}, \change{\mbox{$|S|\!=\!1019$}, and \mbox{$|\overline{S}|\!=\!981$}.  The motivation behind the choice of \mbox{$|V|\!=\!2000$} was to make the size of each subsample approximately 1000 galaxies, meaning that the subsample sizes were sufficiently large  to avoid noisy correlation functions but not too large that the computational complexity made the analysis intractable.}  For our random catalog, we generated a 1 square degree Poisson sample consisting of 200,000 galaxies.

All bins for cross-correlation and autocorrelation were given equal weights in the cost function according to the formalism derived in Section \ref{sec:formalism}.  Furthermore, we chose to use binning schemes with an equal number of galaxy pairs per bin across all bins.  With this choice, the Poisson noise was equal across all bins, unlike binning schemes set by uniform increments in angular separation.

\change{To separate the physics from the mathematics and} test the effectiveness of the optimization itself, we fed Gurobi ground-truth target values by pre-computing all $\alpha_i$'s and $\gamma_i$'s using the real partition of $V$ into $S$ and $\overline{S}$; we also fixed $|S|$ and $|\overline{S}|$ according to equation \ref{eq:ssize} using the real partition of V. Given the ground-truth as target values, we then tested
\begin{enumerate}
\item the time required for Gurobi to complete the optimization
\item the fraction of galaxies assigned correctly
\end{enumerate}
Accordingly, these are the main metrics by which we compare different optimization runs in this section.

In the limit of sufficiently few pairs per bin, the optimization is computationally feasible and recovers the ground-truth solution exactly.  For instance, with a binning scheme of $5$ pairs per bin and $2000$ bins, the optimization recovers the ground-truth partition in $5118$ seconds.  
However, in this case of only 5 pairs per bin, the optimization almost certainly couples to the noise of the ground-truth correlation function target values. This motivated an analysis of a binning scheme of 100 pairs per bin, which improves upon the Poisson noise per bin while still retaining resolution in the autocorrelation signal at short scales by having sufficiently narrow bins.

\change{With 100 pairs per bin and of order 10 bins, the optimization completes in minutes but behaves no better than random assignment in terms of assigning the galaxies correctly.  This can be attributed to the fact that there is a high degree of degeneracy of optimal solutions with 0 cost, given so few bins.  Increasing the number of bins leads to a rapid growth in runtime, during which the optimization galaxy assignment still performs no better than random.  This sharp growth in runtime continues as a function of bin number until approximately 110 bins, after which the optimization does not complete within $10^5$ seconds, a uniformly adopted time cutoff for our computational budget.  Even with 1000 bins, the same phenomenon is observed, and optimization does not complete within the allotted amount of time. 
\iffalse
This transition is so rapid that there does not exist a situation in which the optimization completes within the time cutoff and recovers a solution that performs better than random assignment. \fi
}

In order to reduce computational load and make the optimization feasible, we fixed a fraction of the total galaxy population before beginning the optimization by adding extra constraints to our model that \mbox{$x_u\!=\!1$} if \mbox{$u\!\in\!S$} and \mbox{$x_u\!=\!0$} if \mbox{$u\!\in \!\overline{S}$} for a fraction of galaxies \mbox{$u\!\in\!V$}; including a fixed sample is reasonable because a spectroscopic sample could be used as a fixed population in an application of this method to real data.  In order to explore the optimization with these settings, we ran a series of tests using Gurobi with $80\%$ of galaxies fixed and a binning scheme of 100 pairs per bin. \secondchange{The results are shown in Figure~\ref{fig:100leftright}}. 

%\change{These results are plotted as the red graphs with `+' marks in Figure~\ref{fig:100leftright}}.  In the left plot in Figure \ref{fig:100leftright}, we present the time required for the optimization to finish and the fraction of unfixed galaxies assigned correctly as a \change{function of the total number of pairs of data points used, which is simply 100 times the number of bins.}  In the right plot in Figure \ref{fig:100leftright}, we present the analogous plot with the same quantities as a function of maximum angular binning separation. %

\secondchange{Figure~\ref{fig:100leftright} reveals}
 three distinct regimes \change{for a fixed number of pairs per bin}.  The first regime occurs in the limit of few bins, or short maximum angular scales, in which there are many solutions with 0 cost.  Because there are so many optimal solutions, Gurobi is capable of finding one such optimal solution in a short amount of time;  however, for this same reason, this solution behaves only slightly better than random in assigning the unfixed galaxies  correctly. 
It is worth noting that even in this regime, the amount of time required to complete optimization increases exponentially as a function of number of bins. This increase in runtime is expected because increasing the number of bins increases the number of target values and thus decreases the number of solutions with 0 cost.

The second regime occurs in the limit of many bins,
in which the optimization is communicated a sufficient amount of information about the ground-truth solution through the $\alpha_i$'s and $\gamma_i$'s that it is able to find the ground-truth solution in of order a minute. Consequently, Gurobi assigns 100\% of the unfixed galaxies correctly. \change{Interestingly, this transition occurs when the number of unfixed galaxies is equal to the number of constraints, which is given by the number of bins at the transition point multiplied by 2 (for $\alpha_i$ and $\gamma_i).$}

The third regime is the peak in the amount of time required for the optimization to complete, located between the other two regimes.  As seen in \secondchange{Figure \ref{fig:100leftright}}, the full features of the peak cannot be determined due to the exponential growth in runtime.  The points \change{marked by red circles} correspond to optimization runs that were terminated before completion after $10^5$ seconds, the maximum allowed time in our study; the fraction of galaxies correctly assigned corresponds to that of the optimal solution found before termination.  This peak is the result of a trade-off between the other two regimes: there are relatively few solutions with 0 cost, and they cannot be found quickly by Gurobi; however, there are not enough bin target values to constrain the ground-truth solution immediately. 

In Figure \ref{fig:100leftright}, \change{we also present the analogous results} for a binning scheme with 200 pairs per bin.  \change{These results are plotted as  blue graphs with `x' markers.} The same three characteristic regimes are present; however, the second and third regimes begin at shorter maximum angular binning scales for the binning scheme of 100 pairs per bin than for the binning scheme of 200 pairs per bin. This is a consequence of the fact that for a given maximum angular binning scale, the binning scheme of 100 pairs per bin is fed twice the number of $\alpha_i$'s and $\gamma_i$'s than the binning scheme of 200 pairs per bin and thus has twice the information.  

The results presented in Figure \ref{fig:100leftright} are highly sensitive to the fraction of galaxies fixed before optimization.  For example, in the \change{second} regime of many bins, the optimization requires an order of magnitude more time to complete the optimization for 75\% of galaxies fixed than for 80\% of galaxies fixed; this optimization does not complete within $10^5$ seconds for 70\% of galaxies fixed. 

\change{Reducing the total number of galaxies to $1000$, the optimization requires a lower percentage of galaxies to be fixed in order to recover the ground truth solution in the limit of many bins.  Fixing 65\% of galaxies and using 100 edges per bin and 600 bins, the optimization completes in 195 seconds.  Lowering the percentage fixed to 60\%, the optimization completes in 695 seconds.  Thus, reducing the total number of galaxies to $1000$ still allows for $400$ unfixed galaxies to be assigned, confirming that the percentage of fixed galaxies necessary for the optimization to complete within a given time frame is largely dependent on the total number of galaxies in $V$.}

\iffalse
\begin{figure*}
\epsscale{1.17}
\plottwo{f2a.eps}{f2b.eps}

\caption{Same panels as in Figure~\ref{fig:100leftright} but, instead of 100 pairs per bin, here 200 are used for lower resolution but higher statistical accuracy.}
    \label{fig:200leftright}
\end{figure*}
\fi

\section{Applicability}\label{sec:applicability}

The results presented in Section \ref{sec:results} reveal that in the appropriate regimes, the optimization is computationally feasible when ground-truth values are fed in as the target values.  In any real application of this method, the ground-truth values of the $\alpha_i$'s and $\gamma_i$'s would only be known approximately, and fiducial values are good approximations in the case of many pairs in the bins.
The determination of the appropriate fiducial values is a question of physics rather than of linear programming, and because we have chosen to test only the optimization itself in this paper, so far we have omitted the exploration of the effects of inputting fiducial values on the optimization.  

\subsection{Toward Fiducial Correlations}\label{sec:toward_fiducial}
\begin{figure}
\epsscale{1}
\plotone{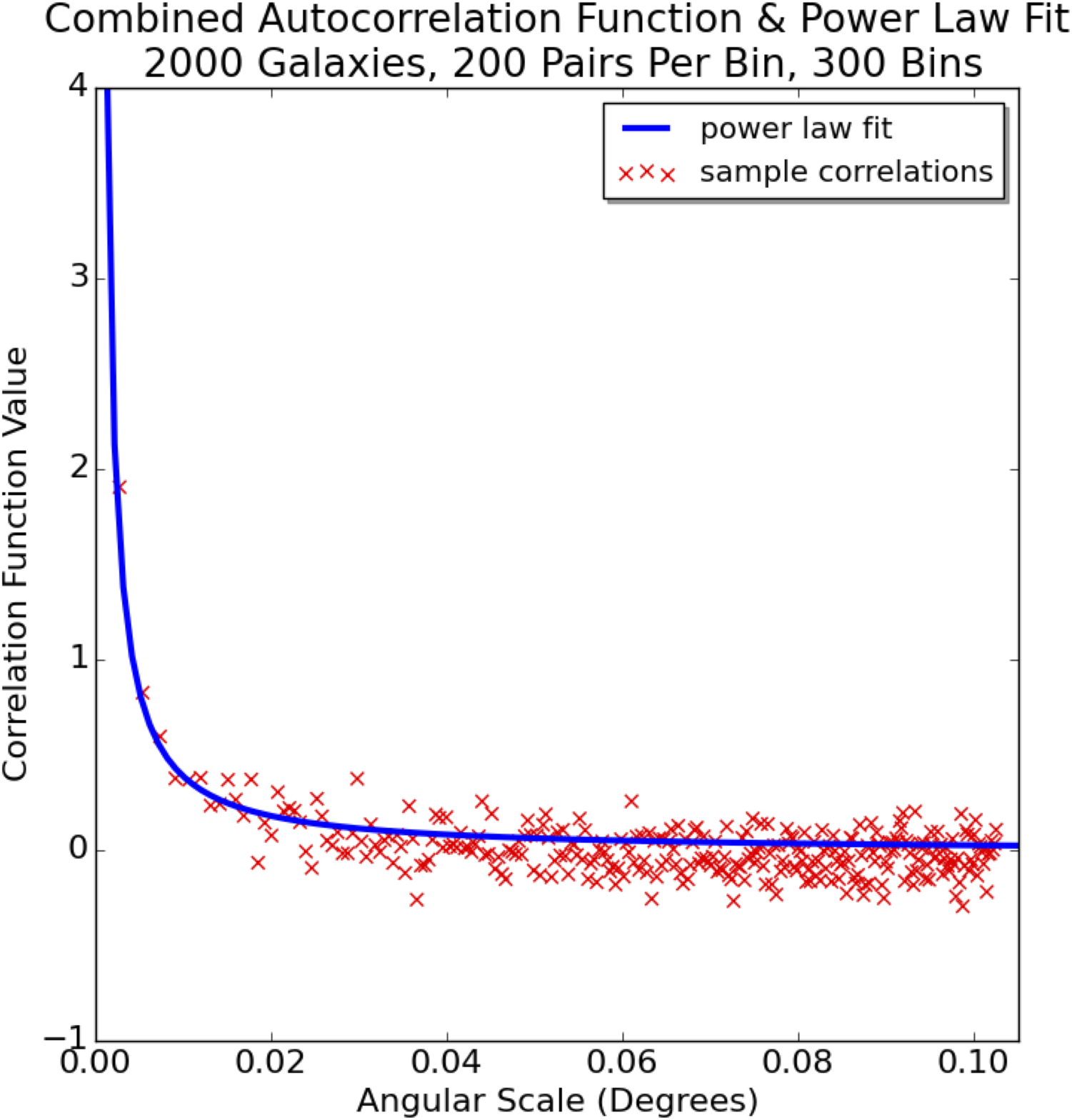}

\caption{A plot of the combined autocorrelation function and power-law fit for 2000 galaxies, 200 pairs per bin, and 300 bins.  Note that  this plot shows the noise in each bin used in the optimization. }
    \label{fig:corrfn}
\end{figure}

\begin{figure}
\epsscale{1}
\plotone{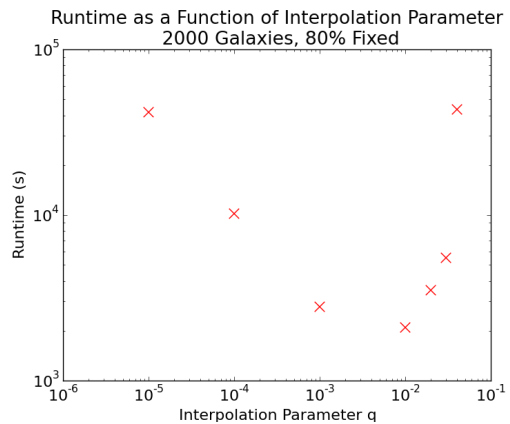}

\caption{A plot of optimization runtime as a function of the interpolation parameter $q$ for 2000 galaxies, 80\% fixed, and a binning scheme of 200 pairs per bin and 300 bins.  For $q = 0$, the runtime is 44 seconds. }
    \label{fig:interpolation}
\end{figure}

%
\iffalse
{\color{blue}{A direct input of {\color{red}{physically-derived fiducial values}} into the optimization causes the optimization not to converge within the allotted amount of time.  In order to explore this further, we test the optimization's response to target values that deviate incrementally from the ground-truth target vales.
}}
\fi
We test the optimization's response to inexact target values by perturbing the $\alpha_i$'s toward values taken from a power law fit of the combined autocorrelation function and perturbing the $\gamma_i$'s toward $0$, the expected cross-correlation.  We accomplish this by using interpolation and setting the target values by varying the interpolation parameter $q \in [0, 1]$ according to the following equations:
\begin{equation}
\alpha_i = w_{m}^{\textrm{auto}}(\theta_i) + q\,\big[w_{pl}^{\textrm{auto}}(\theta_i) - w_{m}^{\textrm{auto}}(\theta_i) \big]
\end{equation}
and 
\begin{equation}
\gamma_i = (1 - q) \, w_{m}^{X}(\theta_i) 
\end{equation}
where $w_{m}^{\textrm{auto}}(\theta_i)$ is the ground-truth combined autocorrelation value in bin $i$, $w_{m}^{X}(\theta_i)$ is the ground-truth cross-correlation value in bin $i$, and $w_{pl}^{\textrm{auto}}(\theta_i)$ is the value of the power-law fit in bin $i$. 
\change{Instead of using the original correlation function 
of the Cox process, 
\mbox{$\xi(r)\!=\!1.59/r^2\!-\!1/r$} \citep{heinis},
we adopted the power-law function which is a good fit at the observed separations,}
as seen in Figure \ref{fig:corrfn}.
%, we present the power law fit of the combined autocorrelation function, as well as the ground-truth values of the autocorrelation function.
This power law fit was used only for the purposes of perturbing the ground-truth combined autocorrelation function by small amounts, as seen in Figure \ref{fig:interpolation}.  \change{In this regard, using a fit versus the actual correlation function of the simulation is inconsequential; here, we are simply exploring slight perturbations from the ground truth values.}
In Figure \ref{fig:interpolation}, we present the time required for the optimization to complete when varying $q$ up to 0.04 for 80\% of galaxies fixed, 200 pairs per bin, and 300 bins.   By varying $q$ only by small amounts, we lessen the impact of an incorrect power law fit.  Furthermore, these binning settings were chosen because the optimization recovered the exact ground-truth partition in 44 seconds given these settings and ground-truth target values.  For all runs in this figure, the optimization exactly recovered the ground-truth solution.

The fact that the optimization still completes and recovers the real partition when given inexact target values for \mbox{$q\!\le\!0.04$}  \change{indicates that the optimization can complete for inexact target correlation function values. However,} for \mbox{$q\!=\!0.05$}, the optimization does not complete within $10^5$ seconds, \change{suggesting a potential limitation to this method in its current formulation.}

The observed phenomenon of an increase in runtime with $q$ is possibly due to the fact that the model might not have a perfect solution and the cost function could become shallow, which leads to a large number of ``nearly optimal" solutions. These solutions must be pruned by the solver to find the global minimum conclusively. However, the pruning methodologies of IP solvers often need to spend a significant amount of time to discard \secondchange{\textit{all}} the ``nearly optimal" solutions, even though they have arrived at a stage of the optimization where all the solutions that are being considered have values that are very close to the true optimal value. To finally discover the true optimal value by weeding through the large number of ``nearly optimal'' ones occupies the bulk of the time for the solver in such situations (see Chapter 2 of~\cite{conforti-ip} for a discussion of these pruning strategies for integer linear programming).

By interpolating a power law fit, we have in effect pulled the $\alpha_i$ target values to a slightly less-noisy correlation function.   In order to resolve fully the question of whether the optimization is finding the ground-truth solution by coupling to noise, $q$ would have to be driven closer to 1; of course, in doing so, the target values would become more dependent on the fiducial values, and physically-correct fiducial values would have to be chosen, \change{ideally via a reproducible method for various galaxy samples}.  However, a visual representation of the noise in each bin of the ground-truth combined autocorrelation function is nonetheless instructive and can be seen in Figure \ref{fig:corrfn}.  

\subsection{Toward Narrow Redshift Slices}\label{sec:toward}
Our ultimate goal is to use this procedure in conjunction with photometric redshift results to sort galaxies into thin redshift slices using only photometric data. The true redshift distribution of photo-z bins will have overlapping tails due to the uncertainties in the estimation. Starting with designations from the photo-z catalog, one can apply our procedure to create a sharper boundary between the bins. Repeating this procedure will yield much improved redshifts.  \change{The sizes of the mock datasets used in Section \ref{sec:results} are reasonable for real application because the optimization could be run iteratively on a large galaxy sample, while keeping the variance in the correlation function estimates reasonably small}.

\change{
An application of immediate relevance is galaxies with degenerate photometric redshifts; sources with specific observed colors may have the probability of being at  multiple redshifts, as explored in \citet{rahmanc}. In many of these cases, there is no additional information in the photometry that can further isolate these sources in redshift. However, there are often other sets of galaxies with narrow photometric redshifts uniquely at the redshift modes of the degenerate set, which may be due to the difference in SEDs of blue and red galaxies, for instance.
\iffalse
%
{\color{magenta}{The clustering of the degenerate sources with the narrow photo-$z$ samples contains information that can be used to sort the former sample into those associated with each of the latter.}}
%
\fi
Consequently, we can leverage the substantial numbers of galaxies with unique photometric redshifts (the fixed galaxies) to break the degeneracy of sources with broad photometric redshift distributions (the unfixed galaxies). In such a case, it is very likely that the fraction of fixed photometric redshift-based galaxies would be similar to those tested in Section~\ref{sec:results}.} 
%
%\textbf{{\color{red}{[add in a description Mubdi's application and describe why 80\% fixed may be completely reasonable for this application]  Tam\'as, I am not sure if we should change the next sentence, given this new application, or leave it be.}}}
%{\color{red}{ Naturally, the fraction of fixed galaxies used in Section \ref{sec:results} would be replaced with spectroscopic samples, although they would represent lower fractions of fixed galaxies than tested in Section \ref{sec:results}. }} 

\section{Conclusions and Future Directions}

We have presented a novel integer linear programming method that enables a galaxy sample to be partitioned into two subsamples such that the angular two-point cross- and autocorrelations of the subsamples are optimized to pre-determined target values. Our approach is the first application of integer linear programming to  \change{astronomical analysis}, which is expected to find other applications as we explore large statistical samples. 
We tested this optimization method using mock catalogs and Gurobi, an optimization solver, and verified that this optimization technique is not only feasible in certain regimes but also provides good solutions, \change{provided that an appropriate percentage of galaxies are fixed before beginning the optimization}. This \change{feasibility} is due to the formulation of the problem using only linear equations.
We explored the applicability of this method and have described how it could be used in the future to estimate the redshifts of individual galaxies using only their celestial coordinates.

Evidently, much more remains to be explored with the applicability of this method, a significant portion of which relates to fiducial values.  For instance, the question of the extent to which the optimization is coupling to noise in the correlation function can only be resolved by analyzing runs with fiducial target values.  Furthermore, much would be learned from re-generating Figure \ref{fig:100leftright} using fiducial values as target values once the best method for selecting fiducial values is identified.  Other variables such as the optimal binning scheme and maximum angular binning separation for the autocorrelation and cross-correlation when using fiducial values would also have to be explored. Ideally, for real application, the number of galaxies per sample could be pushed higher, and the fraction of galaxies fixed could be pushed lower; \change{however, as described in section \ref{sec:toward}, there is at least one application for which this method would be applicable given the fraction of galaxies fixed used throughout section \ref{sec:results}.
Furthermore,} given the rapid improvement in linear programming algorithms and the increase in computing power, these limitations may very well resolve themselves with time. 

Additions to the outlined method could be explored in order to improve runtimes and accuracy. Some of these are mathematical and others physical issues.

(1) In a real application of this method, the optimization could be terminated when the relative gap between the upper and lower bounds of the optimization is below a threshold value, instead of forcing this gap to reach zero in order for the optimization to complete and find the true {\em mathematically} optimal solution. In this case, the solver may report a ``nearly optimal'' solution of the optimization model, as opposed to the true {\em mathematical} optimum of the model. Nevertheless, we feel that from a physical standpoint, the ``nearly optimal'' solutions may be more meaningful than the mathematical optimal which could be influenced by noise. Such a strategy would also make sense when we use fiducial values instead of values from a simulation and would help to deal with the explosion in running time as discussed in Section~\ref{sec:toward_fiducial}.
    
(2) Other than through the $\alpha_i$'s, the power law nature of the autocorrelation function is never explicitly leveraged; it could potentially be exploited in a greedy algorithm, for example, by fixing pairs of shortest angular separations to the same subsample before beginning optimization.  The autocorrelation signal could also potentially be leveraged to a greater extent by using a binning scheme set by uniform increments in angular binning separation, which could provide more bins at the shortest angular scales.
    
(3) It is possible that subsampling the pairs in each bin could decrease runtime without sacrificing accuracy. Considering that the true underlying variables are the classes to which the objects belong ($x_u$), a subset of the pairs ($y_{uv}$) could provide enough constraints at a reduced computational cost. The sampling, however, will probably have to be carefully constructed to optimize performance.

\section*{Acknowledgements}

The authors would like to thank S\'ebastien Heinis for providing mock catalogs and Brice M\'enard for helpful discussions. 
Lance Joseph helped with computational resources.  
\change{The authors also thank the anonymous reviewer for the thorough and insightful report.}
B.L. was supported by the 2015 Herchel Smith-Harvard Undergraduate Science Research Fellowship. 
A.B. gratefully acknowledges partial support from NSF grant CMMI1452820.

\appendix
We formalize the independent autocorrelation method for the Natural Estimator and provide the formalism for the Landy-Szalay Estimator. The mathematical notation for these are somewhat more complicated but the complexity of the algorithm does not increase. \\

\section{Independent Autocorrelation Method for the Natural Estimator}\label{sec:independent}

Here, we introduce the formalism for autocorrelation optimization with the natural estimator in which target values $\alpha_i$ and $\beta_i$ can be set independently, as opposed to the combined autocorrelation method described in \ref{sec:combined}.  The independence of $\alpha_i$ and $\beta_i$ comes at the expense of model complexity:  we must introduce the variables $r_{uv}$ and $\overline{r}_{uv}$ and associated constraints for each unordered pair \mbox{$u,\!v\!\in\!V$}, where neither $r_{uv}$ nor $\overline{r}_{uv}$ can be expressed in terms of  $y_{uv}$.  Therefore, this method effectively triples the number of variables in our  model in comparison to the combined autocorrelation model.  We designate the cost function equivalent of the autocorrelation of $S$ and $\overline{S}$ using the natural estimator as $f_{S}(S)$ and $f_{ \overline{S}}(S)$, respectively. 

In order to formalize the autocorrelation of $S$, we introduce the variable $r_{uv}$.  For each unordered pair \mbox{$u,\!v\!\in\!V$}, we define $r_{uv}$ as follows:
\begin{equation}
   r_{uv} = r(u,v) = \left\{
     \begin{array}{ll}
       1 & : u, v \in S\\
       0 & : \text{otherwise}
     \end{array}
   \right.
\end{equation}

In addition, to formalize the autocorrelation of $\overline{S}$, we introduce the analogous variable $\overline{r}_{uv}$.  For each unordered pair \mbox{$u,\!v\!\in\!V$}, we define $\overline{r}_{uv}$ as follows:
\begin{equation}
   \overline{r}_{uv} = \overline{r}(u,v) = \left\{
     \begin{array}{ll}
       1 & : u, v \in \overline{S}\\
       0 & : \text{otherwise}
     \end{array}
   \right.
\end{equation} 

Just as with $y$'s, we must add constraints to relate $x$'s to $r$'s and $\overline{r}$'s:

\begin{equation}
  \begin{split}
r_{uv} \ge x_u + x_v - 1
  \qquad
&r_{uv} \le x_u \\
r_{uv} \le x_v
\qquad
&r_{uv} \ge 0
  \end{split}
\end{equation}

\begin{equation}
  \begin{split}
\overline{r}_{uv} \ge 1 - x_u - x_v
  \qquad
&\overline{r}_{uv} \le 1 - x_u \\
\overline{r}_{uv} \le 1 - x_v
\qquad
&\overline{r}_{uv} \ge 0
  \end{split}
\end{equation}
For the autocorrelation of $S$, we minimize:
\begin{equation}\label{eq:as}
\sum_{i = 1}^n  \bigg| \bigg( a^S_i \sum\limits_{(u, v) \in VV_i} r_{uv} \bigg) - (1 + \alpha_i) \bigg|,
\quad
a^S_i= \frac{|R|(|R| - 1)}{|RR_i|} \cdot \frac{1}{|S|(|S| - 1)}
\end{equation}
and for the autocorrelation of $\overline{S}$, we minimize:
\begin{equation}\label{eq:asbar}
\sum_{i = 1}^n  \bigg| \bigg( a^{\overline{S}}_i \sum\limits_{(u, v) \in VV_i} \overline{r}_{uv} \bigg) - (1 + \beta_i) \bigg|,
\quad
a^{\overline{S}}_i= \frac{|R|(|R| - 1)}{|RR_i|} \cdot \frac{1}{|\overline{S}|(|\overline{S}| - 1)}
\end{equation}

\iffalse
Thus, $f_{N, S}(S)$ takes the final form of:
\begin{equation}
f_{N, S}(S) = \sum_{i = 1}^n \psi^{S}_i
\end{equation}
where for each $\psi_i$, we add the following two constraints:
\begin{equation}
\begin{split}
\bigg( w^S_i \sum\limits_{u, v \in VV_i} r_{uv} \bigg) - (1 + \alpha_i) &\le \psi^{S}_i\\
\bigg( w^S_i \sum\limits_{u, v \in VV_i} r_{uv} \bigg) - (1 + \alpha_i) &\le - \psi^{S}_i
  \end{split}
\end{equation}
and $f_{N, \overline{S}}(S)$ takes the final form of:
\begin{equation}
f_{N, \overline{S}}(S) = \sum_{i = 1}^n \psi^{\overline{S}}_i
\end{equation}
where for each $\psi_i$, we add the following two constraints:
\begin{equation}
\begin{split}
\bigg( w^{\overline{S}}_i \sum_{u, v \in VV_i} \overline{r}_{uv} \bigg) - (1 + \beta_i) &\le \psi^{\overline{S}}_i\\
\bigg( w^{\overline{S}}_i \sum_{u, v \in VV_i} \overline{r}_{uv} \bigg) - (1 + \beta_i) &\le - \psi^{\overline{S}}_i
  \end{split}
\end{equation}
\fi

In order to convert $f_{S}(S)$ and $f_{\overline{S}}(S)$ into their final forms, we must eliminate the absolute values using the method described in Sections~\ref{sec:cross} and \ref{sec:combined}.  In this formalization using independent autocorrelations for $S$ and $\overline{S}$ and using the natural estimator, the full model consists of the cost function:
\begin{equation}
f(S) = f_{X}(S) + f_{S}(S) + f_{ \overline{S}}(S)
\end{equation}
and all of the associated constraints.

\section{The Landy-Szalay Estimator}\label{sec:LS}
Here, we introduce the formalism for optimization with the Landy-Szalay estimator.  Using the Landy-Szalay estimator, the autocorrelation and cross-correlation function estimates of two samples $D$ and $D'$ in bin $i$ are given by:
\begin{equation}\label{eq:ls-cross}
\fn_{ X, i} = \frac{|R|(|R| - 1)}{2 \cdot |RR_i|} \cdot \frac{|DD'_i|}{|D||D'|} - \frac{|R| - 1}{ 2 \cdot|RR_i|} \Bigg( \frac{|D R_i| }{|D|} + \frac{|D' R_i|}{|D'|} \Bigg) + 1
\end{equation}
\begin{equation}\label{eq:ls-D}
\fn_{D, i} = \frac{|R|(|R| - 1)}{|RR_i|} \cdot \frac{|DD_i|}{|D|(|D| - 1)} - \frac{|R| - 1}{|RR_i|} \cdot \frac{|D R_i| }{|D|}  + 1
\end{equation}
\begin{equation}\label{eq:ls-D'}
\fn_{D', i} = \frac{|R|(|R| - 1)}{|RR_i|} \cdot \frac{|D'D'_i|}{|D'|(|D'| - 1)} - \frac{|R| - 1}{|RR_i|} \cdot \frac{|D' R_i|}{|D'|}  + 1
\end{equation}
In building the ILP model for the natural estimator, we have already modeled the first terms in all three expressions~\eqref{eq:ls-cross},~\eqref{eq:ls-D}, and \eqref{eq:ls-D'}.  Thus, we only need to translate the second terms in these expressions.  Fortunately, they can be expressed entirely in terms of constants and $x_u$'s.

We begin with the $SR$ term. In \eqref{eq:ls-cross} and \eqref{eq:ls-D}, this term in the $i$th bin is defined as:
\begin{equation}
\frac{|R| - 1}{2 \cdot |RR_i|} \cdot \frac{|S R_i|}{|S|} 
\end{equation}
Furthermore, we know that $|S R_i|$ is defined as the number of unordered pairs \mbox{$u,\!v\!\in\!SR$} such that the angular separation between $u$ and $v$ lies in bin $i$.  Defining $|u R_i|$ for a given \mbox{$u\!\in\!S$} as the number of ordered pairs \mbox{$u,\!r$} such that \mbox{$r\!\in\!R$} and the angular separation between $u$ and $r$ lies in bin $i$, we can re-express $|SR_i|$:
\begin{equation}
|SR_i| = \sum_{u \in S} |u R_i|
\end{equation}
We can in turn express this in terms of our variables $x_u$:
\begin{equation}
|SR_i| = \sum_{u \in V} |u R_i| x_u 
\end{equation}

By generalizing to summing over all galaxies in $V$ as opposed to just galaxies in $S$, we have eliminated any dependence on the partitioning of $V$ except in the variables themselves.  Furthermore, for a given galaxy \mbox{$u\! \in\!V$}, $|u R_i|$ is a constant that can be pre-computed before optimization.  We now define the weight $\mu^{S}_i$ to absorb all of these constants:
\begin{equation}
\mu^{S}_{i, u}  =  \frac{|R| - 1}{2 \cdot |RR_i|} \cdot \frac{|uR_i| }{|S|}
\end{equation}
Thus, in the $i$th bin, the term involving $SR$ becomes:
\begin{equation}
\sum_{u \in V} \mu^{S}_{i, u} x_u
\end{equation}
We can define the term involving $\overline{S} R$ in the $i$th bin analogously:
\begin{equation}
\sum_{u \in V} \mu^{\overline{S}}_{i, u} \big(1 - x_u \big)
\end{equation}
where:
\begin{equation}
\mu^{\overline{S}}_{i, u}  =  \frac{|R| - 1}{2 \cdot |RR_i|} \cdot \frac{|uR_i| }{|\overline{S}|}
\end{equation}
Thus, referring to \eqref{eq:ls-cross}, \eqref{eq:ls-D}, \eqref{eq:ls-D'}, we can now express $\fn_{X}$, $\fn_{S}$, and $\fn_{\overline{S}}$, respectively, in terms of our binary variables:
\begin{equation}
\begin{array}{l}
\fn_{X, i} =  \left( a_i \sum\limits_{(u, v) \in VV_i}   y_{u v} \right) -\sum_{u \in V} \mu^{S}_{i, u} x_u - \sum\limits_{u \in V} \mu^{\overline{S}}_{i, u} \big(1 - x_u \big) + 1 \\
\fn_{S, i} =  \left( a^S_i \sum\limits_{(u, v) \in VV_i} r_{u v} \right) - 2 \sum_{u \in V} \mu^{S}_{i, u} x_u + 1
\\
\fn_{\overline{S}, i} =  \left( a^{\overline{S}}_i \sum\limits_{(u, v) \in VV_i} \overline{r}_{u v} \right) - 2 \sum_{u \in V} \mu^{\overline{S}}_{i, u} \big(1 - x_u \big) + 1
\end{array}
\end{equation}
where $a_i$ has been defined in equation \ref{eq:a}, and $a^S_i$ and $a^{\overline{S}}_i$ have been defined in equations \ref{eq:as} and \ref{eq:asbar}, respectively.

The cost function equivalents of $\fn_{X}$, $\fn_{S}$, and $\fn_{\overline{S}}$, given by $f_{X}(S)$, $f_{S}(S)$, and $f_{ \overline{S}}(S)$, respectively, can now be converted to their final forms by eliminating the absolute values using the method described in Sections~\ref{sec:cross} and \ref{sec:combined}.  Thus, our full model consists of the cost function:
\begin{equation}
f(S) = f_{X}(S) + f_{S}(S) + f_{\overline{S}}(S)
\end{equation}
and all associated constraints.\\


\begin{thebibliography}{}
\bibitem[Ben\'itez(2000)]{benitez00}
Ben\'itez, N. 2000, 
ApJ, 536, 571
\bibitem[Benjamin et~al.(2010)]{benjamin10} Benjamin, J., van Waerbeke, L., M\'enard, B., \& Kilbinger, M. 2010, MNRAS,
408, 1168
\bibitem[Brammer et~al.(2008)]{brammer08}
Brammer, G., van Dokkum, P., \& Coppi, P.
2008, ApJ, 686, 1503
\bibitem[Budav\'ari et~al.(2000)]{budavari00}
Budav\'ari, T., Szalay, A., Connolly, A., Csabai, I., \& Dickinson, M.
2000, AJ, 120, 1588
\bibitem[Budav\'ari et~al.(2001)]{budavari01}
Budav\'ari, T., Csabai, I., Szalay, A., Connolly, A., \& Szokoly, G.
2001, AJ, 122, 1163
\bibitem[Budav\'ari(2008)]{budavari08}
Budav\'ari, T.
2008, ApJ, 695, 747
\bibitem[Conforti et~al.(2015)]{conforti-ip} Conforti, M., Cornu\'ejols, G., Zambelli, G., Integer Programming, {\em Graduate Texts in Mathematics, Springer-Verlag}, 2015.
\bibitem[Connolly et~al.(1995)]{connolly95}
Connolly, A., Csabai, I., Szalay, A., Koo, D., Kron, R., \& Munn, J.
1995, AJ, 110, 2655
\bibitem[Feldmann et~al.(2006)]{feldmann06} Feldmann, R., et~al.
2006, MNRAS, 372, 565
\bibitem[Gurobi(2015)]{gurobi} Gurobi Optimization, Inc., 2015, Gurobi Optimizer Reference Manual,  http://www.gurobi.com
\bibitem[Heinis et~al.(2009)]{heinis} Heinis, S., Budav\'ari, T., \& Szalay, A. 2009, ApJ, 705, 739
\bibitem[Kerscher et~al.(2000)]{kerscher} Kerscher, M., Szapudi, I.,  \& Szalay, A. 2000, ApJ, 535, L13
\bibitem[Koo(1985)]{koo85}
Koo, D.
1985, AJ, 90, 418
\bibitem[Koo(1999)]{koo99} Koo, D. 
1999, in ASP Conf. Ser. 191, Photometric Redshifts and High Redshift Galaxies, ed. Weymann, R.,  Storrie-Lombardi, L., Sawicki, M., \& Brunner, R. (San Francisco, CA: ASP), 3
\bibitem[Landy \& Szalay(1993)]{landy} Landy, S. D., \& Szalay, A. S. 1993, ApJ, 412, 64
\bibitem[Menard(2013)]{menard13} M\'enard, B., Scranton, R., Schmidt, S., Morrison, C., Jeong, D., Budav\'ari, T., \& Rahman, M. 
2013, ArXiv e-prints, arXiv:1303.4722
\bibitem[Newman(2008)]{newman08} Newman, J. 
2008, ApJ, 684, 88
\bibitem[Rahman et~al.(2015a)]{rahmana} Rahman, M., M\'enard, B., \& Scranton, R. 2015a, ArXiv e-prints, arXiv:1508.03046
\bibitem[Rahman et~al.(2015b)]{rahmanb} Rahman, M., M\'enard, B., Scranton, R., Schmidt, S., \& Morrison, C. 
2015b, MNRAS, 447, 3500
\bibitem[Rahman et al.(2016)]{rahmanc} Rahman, M., Mendez, A.~J., M{\'e}nard, B., et al.\ 2016, \mnras, 460, 163 
\bibitem[Schmidt et~al.(2013)]{schmidt2013} Schmidt, S. J., M\'enard, B., Scranton, R., Morrison, C., \& McBride, C. K.
2013, MNRAS, 431, 3307
\bibitem[Schmidt et~al.(2015)]{schmidt2015} Schmidt, S. J., M\'enard, B., Scranton, R., et al. 
2015, MNRAS, 446, 2696
\end{thebibliography}
\end{document}